\documentclass[12pt]{article}
\usepackage[margin=2.5cm]{geometry}
\usepackage{graphics} 
\usepackage{epsfig} 
\usepackage{amsmath} 
\usepackage{amssymb}  
\usepackage{url}

\usepackage{subfigure}
\usepackage{amsmath}
\usepackage{amsthm} 
\usepackage{color}
\usepackage{footnote}
\usepackage{algorithm}
\usepackage{algpseudocode}
\usepackage{algorithmicx}
\usepackage{multirow} 
\usepackage{booktabs}
\usepackage{graphicx}
\usepackage{amssymb}
\usepackage{amsbsy}
\usepackage{array}
\usepackage{longtable}
\usepackage{epstopdf}
\usepackage{pbox}
\usepackage{breqn}
\usepackage{mathrsfs}
\usepackage{multicol}
\usepackage{supertabular}
\usepackage{enumerate}
\usepackage[colorinlistoftodos]{todonotes}
\usepackage{url}
\usepackage[justification=centering]{caption}
\usepackage{pgfplots}
\usepackage{caption}
\newtheorem{thm}{Theorem}

\newtheorem{prp}{Proposition}

\title{\LARGE \bf
Resilience of Dynamic Routing in the Face of Recurrent and Random Sensing Faults
}

\author{Qian Xie and Li Jin
\thanks{This work was in part supported by NYU Tandon School of Engineering and the C2SMART University Transportation Center. The authors appreciate the discussion with Profs. Saurabh Amin and Patrick Jaillet at Massachusetts Institute of Technology.}
\thanks{Q. Xie and L. Jin are with the Department of Civil and Urban Engineering, New York University Tandon School of Engineering, Brooklyn, NY, USA. (emails: qianxie@nyu.edu, lijin@nyu.edu)}
}
\date{}

\pgfplotsset{width=4.35cm,compat=1.14}
\begin{document}
\maketitle
\thispagestyle{plain}
\pagestyle{plain}

\begin{abstract}
Feedback dynamic routing is a commonly used control strategy in transportation systems.
This class of control strategies relies on real-time information about the traffic state in each link.
However, such information may not always be observable due to temporary sensing faults. 
In this article, we consider dynamic routing over two parallel routes, where the sensing on each link is subject to recurrent and random faults. The faults occur and clear according to a finite-state Markov chain. When the sensing is faulty on a link, the traffic state on that link appears to be zero to the controller.
Building on the theories of Markov processes and monotone dynamical systems, we derive lower and upper bounds for the resilience score, i.e. the guaranteed throughput of the network, in the face of sensing faults by establishing stability conditions for the network.
We use these results to study how a variety of key parameters affect the resilience score of the network.
The main conclusions are: (i) Sensing faults can reduce throughput and destabilize a nominally stable network; (ii) A higher failure rate does not necessarily reduce throughput, and there may exist a worst rate that minimizes throughput; (iii) Higher correlation between the failure probabilities of two links leads to greater throughput; (iv) A large difference in capacity between two links can result in a drop in throughput.
\end{abstract}


\section{Introduction}

The rapidly growing deployment of traffic sensing and vehicle-to-vehicle/infrastructure (V2V/ V2I) communications has enabled the concept of intelligent transportation system (ITS). In ITS, system operators and travelers have access to real-time traffic conditions and can thus make better decisions. Dynamic routing is a typical ITS capability, which is conducted via route guidance tools such as Google Maps and WAZE. System operators can also influence routing via tolling and instructions for traffic diversion, which also rely on real-time traffic conditions.
A major challenge for dynamic routing in ITS is how to ensure system functionality and efficiency under a variety of sensing faults. Quality of sensing and communications significantly affects system performance. However, data health is a serious issue that system operators must face. On some highways, up to 30\%-40\% of loop sensors do not report accurate measurements \cite{van2005accurate,rajagopal2008distributed}; similar issue exists for camera sensors. Even though some routing guidance tools may have certain internal fault detection and correction actions, the benefits of such actions can be further evaluated. Moreover, without appropriate fault-tolerant mechanisms, feedback control algorithms may make decisions based on wrong information, and ITS may even perform worse than a comparable conventional transportation system. Therefore, ITS will not be well accepted by the public and transportation authorities unless the impact of sensing faults is adequately evaluated and addressed. However, such impact has not been well understood, and practically relevant fault-tolerant routing algorithms have not been developed.

\begin{figure}[hbt]
    \centering
    \includegraphics[width=8cm]{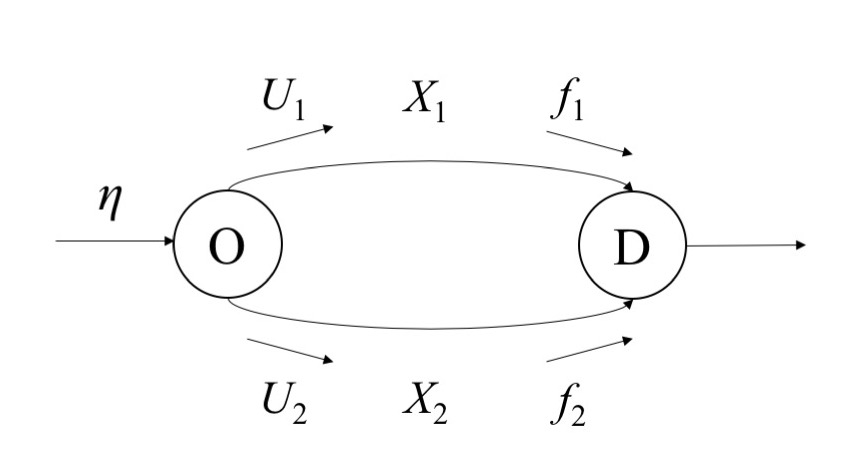}
    \includegraphics[width=6cm]{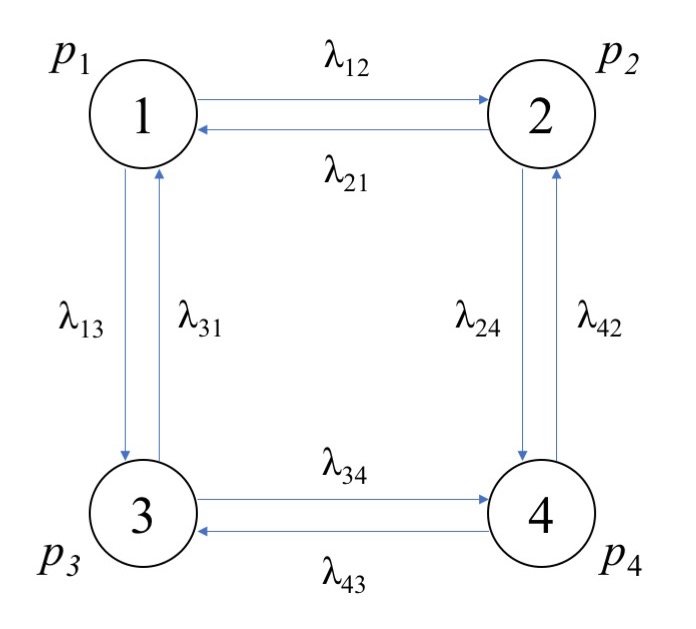}
    \caption{The two-link network and the Markov chain representing network switches among the sensing fault modes.}
    \label{fig:network}
\end{figure}

In this paper, we propose a novel model that synthesizes traffic flow dynamics and stochastic sensing faults. Based on this model, we evaluate the impact of faults on fault-unaware routing algorithm and derive practically relevant insights for designing fault-tolerant routing algorithms in ITS.
We consider the routing problem over two parallel links, as shown in Fig.~\ref{fig:network}. 
Our approach and results can be extended to more complex networks and a broader class of ITS control capabilities, such as ramp metering and speed limit control.
We consider a stochastic model, since in practice it is not easy to deterministically predict when and where a sensing fault will occur.
We will show that this model leads to tractable analysis and insightful results for fault-tolerant design of ITS.
We study the stability and guaranteed throughput of the network, which we consider as the resilience score. We also establish the link between the resilience score and key model parameters, including the number of fault-prone links and the average frequency and duration of faults.

Existing model-based traffic management approaches typically assume complete knowledge of the traffic condition \cite{gomes2006optimal,coogan2015compartmental,reilly2015adjoint,yu2019traffic}, but feedback traffic management for ITS in the face of sensing faults has not been well studied. 
Como et al. \cite{como2012robustI} studied the resilience of distributed routing in the face of physical disruptions to link capacities in a dynamic flow network.
Lygeros et al. \cite{lygeros2000fault} proposed a conceptual framework for fault-tolerant traffic management, but the concrete algorithms are still yet to be developed.
A body of work on fault-tolerant control has been developed for a class of dynamical systems \cite{patton1997fault,blanke2006diagnosis,zhang2008bibliographical}. However, very limited results are available for recurrent and random faults. In addition, there exist some results on adaptive/learning-based fault-tolerant control with applications in electrical/mechanical/aerospace engineering \cite{zhang2004adaptive,mhaskar2006integrated,tang2007adaptive}, but these results are not directly applicable to ITS, nor do they explicitly consider stochastic sensing faults.

Our modeling approach is innovative in that we model the occurrence and clearance of sensing faults as a finite-state, continuous-time Markov process. If the sensing on a link is normal, travelers know the true traffic state (traffic density) on the link. If the sensing is faulty, the traffic state will appear to be zero to the travelers. Besides such denial-of-service, our modeling approach can also be extended to incorporate other forms of sensing faults, such as bias and distortion.
We adopt the classical logit model \cite{ben1985discrete} for routing; the essential principle of this model is that more traffic will go to a less congested link.
When the sensing on a link is faulty, travelers may mistakenly consider a congested link to be uncongested. We show that such faulty information may affect the network's throughput.
The discrete states of the Markov process are essentially modes for the flow dynamics, which govern the evolution of the continuous states.
Hence, our model belongs to a class of stochastic processes called piecewise-deterministic Markov processes \cite{davis1984piecewise, benaim2015qualitative}. Similar models have been used for demand/capacity fluctuations \cite{jin2019analyzing,jin2019behavior}; this paper extends the modeling approach to sensing faults.

A key step for resilience analysis is to determine the stability of the traffic densities under various combinations of parameters.
We study the stability of the network based on the theory of continuous-time Markov processes \cite{meyn1993stability}.
We derive a necessary condition for stability by constructing a positively invariant set for the dynamic flow network.
We derive a sufficient condition by considering a quadratic, switched Lyapunov function that verifies the Foster-Lyapunov drift condition.
We exploit a special property of the flow dynamics, called cooperative dynamics \cite{hirsch1985systems,smith2008monotone}, to derive an easy-to-check stability criterion, which states that the network is stable if there exists a queuing state such that the rate of change of the fastest growing queue averaged over the modes is negative.

Based on the stability analysis, we analyze the network's throughput (resilience score).
We define throughput as the maximal inflow that the network can take while maintaining stable.
As a baseline, we first study the behavior of the network if both links have the same flow functions.
We perturb the baseline in multiple dimensions (probability and correlation of sensing faults on two links) and analyze how throughput can be affected. We also show that throughput reduces as the two link's asymmetry increases.

The main contributions of this paper include (i) a novel stochastic model for sensing fault-prone transportation networks, (ii) easy-to-check stability conditions for the network, and (iii) resilience analysis under various settings.
The rest of this paper is organized as follows.
In Section~\ref{sec:model}, we introduce the dynamic flow model with sensing faults.
In Section~\ref{sec:stability}, we establish the stability conditions.
In Section~\ref{sec:resilience}, we study the resilience score under various scenarios.
In Section~\ref{sec:concluding}, we summarize the conclusions and mention several future directions.
\section{Dynamic flow model with sensing faults}\label{sec:model}

Consider the two-link network in Fig.~\ref{fig:network}.
Let $U_k(t)$ be the flow into link $k\in\{1,2\}$ and $X_k(t)$ be the traffic density of link $k$ at time $t$. The capacity of link $k$ is $F_k\in[0, 1]$ where $F_1+F_2=1$.
The flow out of link $k$ is $f_k(X_k(t))$, which is specified by the flow function
\begin{align}\label{f}
    f_k(x_k)=F_k(1-e^{-x_k}),\quad k=1,2.
\end{align}
The source node is subject to a constant demand $\eta\ge0$, which is considered as a model parameter rather than a state or input variable in the subsequent analysis.

Travelers can observe the state $X(t)$. However, the observation is not always accurate.
We consider the sensing on each link to be stochastically switching between a ``good'' and a ``bad'' mode.
That is, we consider a set $\mathcal S=\{1,2,3,4\}$ of \emph{sensing fault modes}.
The network switches between the two modes according to the Markov chain in Fig.~\ref{fig:network}.
Each mode $s\in\mathcal S$ is characterized by a \emph{fault mapping} $T_s:\mathbb R_{\ge0}^2\to\mathbb R_{\ge0}^2$
\begin{align}
    T_1(x)=\left[\begin{array}{c}
         x_1\\
         x_2 
    \end{array}\right], \ 
    T_2(x)=\left[\begin{array}{c}
         0\\
         x_2 
    \end{array}\right], \ 
    T_3(x)=\left[\begin{array}{c}
         x_1\\
         0 
    \end{array}\right], \ 
    T_4(x)=\left[\begin{array}{c}
         0\\
         0
    \end{array}\right].
\end{align}
In mode $s$, the observed state is 
$$
\hat x=T_s(x).
$$

At the source node, the demand $\eta$ is distributed to each link according to a \emph{routing policy} $\mu:\mathbb R_{\ge0}^2\to\mathbb R_{\ge0}^2$, which specifies the fraction of inflow that goes to each link according to the logit model
\begin{align}\label{muk}
    \mu_k(x)=\frac{e^{-\beta \hat x_k}}{\sum_{j=1}^2e^{-\beta \hat x_j}},\quad k=1,2.
\end{align}
Note that the routing is based on the observed state rather than the true state.

For notational convenience, with a slight abuse of notation, we write
\begin{align}\label{mu}
 \mu(s,x)=\mu(T_s(x)).
\end{align}
That is, the routing policy can be viewed as a switching function $\mu:\mathcal S\times\mathbb R_{\ge0}^2\to[0,1]^2$ with a discrete argument $s\in\mathcal S$ and a continuous argument $x\in\mathbb R_{\ge0}^2$. Finally, we emphasize that we consider $\eta$ as a model parameter rather than a state or input variable in the subsequent analysis.

Then, we define the dynamics of the hybrid-state process $\{(S(t),X(t));t>0\}$ as follows.
The discrete-state process $\{S(t);t>0\}$ of the mode is a time-invariant finite-state Markov process that is independent of the continuous-state process $\{X(t);t>0\}$ of the traffic densities. The state space of the finite-state Markov process is $\mathcal S$. The \emph{transition rate} from mode $s$ to mode $s'$ is $\lambda_{s,s'}$. Without loss of generality, we assume that $\lambda_{s,s}=0$ for all $s\in\mathcal S$ \cite{strang1993introduction}.
Hence, the discrete-state process evolves as follows:
$$
\Pr\{S(t+\delta)=s'|S(t)=s\}=\lambda_{s,s'}\delta+\mathrm o(\delta),
\quad \forall s'\neq s,\ \forall s\in\mathcal S.
$$
where $\delta$ denotes infinitesimal time.
We assume that the discrete-state process is ergodic \cite{gallager2013stochastic} and admits a unique steady-state probability distribution $\{p_s;s\in\mathcal S\}$ satisfying
\begin{subequations}
\begin{align}
    &p_s\sum_{s'\neq s}\lambda_{s,s'}=\sum_{s'\neq s}p_{s'}\lambda_{s',s},
    \quad\forall s\in\mathcal S,\label{pa}\\
    &p_s\ge0, \quad\forall s\in\mathcal S,\\
    &\sum_{s\in\mathcal S}p_s=1.\label{pz}
\end{align}
\end{subequations}

The continuous-state process $\{X(t);t>0\}$ is defined as follows.
For any initial condition $S(0)=s$ and $X(0)=x$,
\begin{align}
\frac d{dt}X_k(t)=\eta\mu_k\Big(S(t),X(t)\Big)-f_k\Big(X(t)\Big),
\quad t\ge0,\ 
k=1,2.
\label{eq_continuous}    
\end{align}

Note that the routing policy $\mu$ defined in \eqref{muk}-\eqref{mu} and the flow function $f$ defined in \eqref{f} ensure that $X(t)$ is continuous in $t$. We can define the flow dynamics with a vector field $G:\mathcal S\times\mathbb R_{\ge0}^2\to\mathbb R^2$ as follows:
\begin{align}
    G(s,x):=\eta\mu(s,x)-f(x).
    \label{eq_G}
\end{align}

The joint evolution of $S(t)$ and $X(t)$ is in fact a piecewise-deterministic Markov process and can be described compactly using an infinitesimal generator \cite{davis1984piecewise, benaim2015qualitative}
\begin{align*}
\mathcal Lg(s,x)=
\Big(\eta\mu(s,x)-f(x)\Big)^T\nabla_xg(s,x) 
+
\sum_{s'\in\mathcal S}\lambda_{s,s'}(g(s',x)-g(s,x)).
\end{align*}
for any differentiable function $g$.

The network is stable if there exists $Z<\infty$ such that for any initial condition $(s,x)\in\mathcal S\times\mathbb R_{\ge0}^2$
\begin{align}\label{eq_bounded}
  \limsup_{t\to\infty}\frac1t\int_{r=0}^t\mathrm E[|X(r)|]dr\le Z.
\end{align}
This notion of stability follows a classical definition \cite{dai1995stability}, some authors name it as ``first-moment stable'' \cite{shi2015survey}.
The rest of this paper is devoted to establishing and analyzing the relation between the stability of the continuous-state process $\{X(t);t>0\}$ and the demand $\eta$.
\section{Stability analysis}\label{sec:stability}

The main result of this section is as follows.
\begin{thm}\label{thm_stability}
Consider two parallel links with sensors switching between two modes as defined in section \ref{sec:model}.
\begin{enumerate}
    \item A necessary condition for stability is that
    \begin{subequations}
\begin{align}
    &\eta\Big(\frac{1}{e^{-\beta \underline x_2}+1}p_2+\frac12p_4\Big)\le F_1,\label{necessary1}\\
    &\eta\Big(\frac{1}{e^{-\beta \underline x_1}+1}p_3+\frac12p_4\Big)\le F_2,\label{necessary2}\\
    &\eta<1.\label{necessary3}
\end{align}
\end{subequations}
where $\underline x_k$ is the solution to
\begin{align*}
    \eta\frac{e^{-\beta\underline x_k}}{1+e^{-\beta\underline x_k}}=F_k(1-e^{-\underline x_k})
\end{align*}
for $k=1,2$.
    \item A sufficient condition for stability is that
    there exists $\theta\in\mathbb R_{\ge0}^2$ such that
\begin{align}
    \sum_{s=1}^4p_s\max_{k\in\{1,2\}}\Big\{\eta\frac{e^{-\beta T_{s,k}(\theta_k)}}{e^{-\beta T_{s,k}(\theta_2)}+e^{-\beta T_{s,k}(\theta_1)}}
    -F_k(1-e^{-\theta_k})\Big\}<0\label{sufficient}
\end{align}
\end{enumerate}
\end{thm}

The rest of this section is devoted to the proof of the above result.

\subsection{Proof of necessary condition}
An apparent necessary condition for stability is
\begin{align}\label{eq_KF}
    \eta<1.
\end{align}
If this does not hold, then the network is unstable even in the absence of sensing faults \cite{jin2018stability}.


First, an invariant set of the process $\{X(t);t>0\}$ is $\mathcal M=[\underline x_1,\infty)\times[\underline x_2,\infty)$. To see this, note that for any $s\in\mathcal{S}$ and for any $(x_1, x_2)$ such that $(x_1, x_2)\notin\mathcal{M}$, the vector $G$ of time derivatives of the traffic densities has a non-zero component that points to the interior of the invariant set $\mathcal{M}$; see Figure \ref{fig:quiver}.

\usepgfplotslibrary{fillbetween}
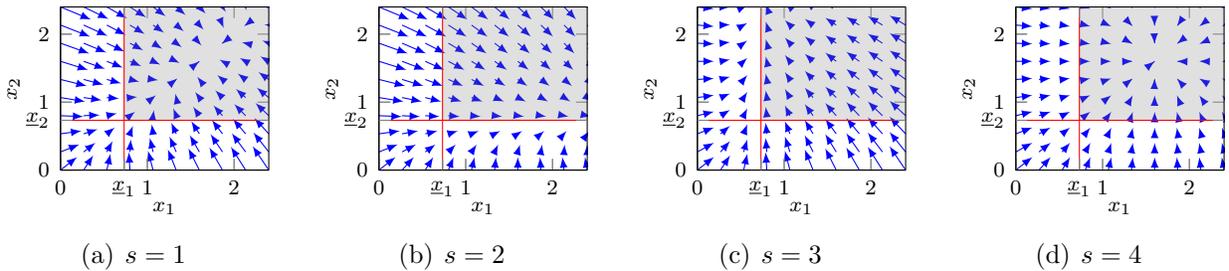
\begin{figure}[h]
    \centering
    \subfigure [$s=1$] {
        \begin{tikzpicture}[font=\scriptsize]
          \begin{axis}[
              view     = {0}{90},
              domain   = 0:2.4,
              y domain = 0:2.4,
              samples  = 10,
              xmax     = 2.4,
              ymax     = 2.4,
              extra x ticks = {0.732668},
              extra x tick labels={$\underline{x}_1$},
              extra y ticks = {0.732668},
              extra y tick labels={$\underline{x}_2$},
              xlabel   = $x_1$,
              ylabel   = $x_2$,
              xlabel style={yshift=0.2cm},
              ylabel style={yshift=-0.2cm}
            ]
            \addplot3 [blue, quiver={u={0.8*e^(-x)/(e^(-x)+e^(-y))-0.5*(1-e^(-x))}, v={0.8*e^(-y)/(e^(-x)+e^(-y))-0.5*(1-e^(-y))}, scale arrows=0.48,
                       every arrow/.append style={-latex}}] (x,y,0);  
            \draw [very thin, draw=gray, fill=gray, opacity=0.24]
               (0.732668,0.732668) rectangle (2.4,2.4) ;
            \draw[red] (0.732668,2.4) -- (0.732668,0);
            \draw[red] (2.4,0.732668) -- (0,0.732668);
          \end{axis}
        \end{tikzpicture}
    }
    \hspace{0.1cm}
    \subfigure [$s=2$] {
        \begin{tikzpicture}[font=\scriptsize]
          \begin{axis}[
              view     = {0}{90},
              domain   = 0:2.4,
              y domain = 0:2.4,
              samples  = 10,
              xmax     = 2.4,
              ymax     = 2.4,
              extra x ticks = {0.732668},
              extra x tick labels={$\underline{x}_1$},
              extra y ticks = {0.732668},
              extra y tick labels={$\underline{x}_2$},
              xlabel   = $x_1$,
              ylabel   = $x_2$,
              xlabel style={yshift=0.2cm},
              ylabel style={yshift=-0.2cm}
            ]
            \addplot3 [blue, quiver={u={0.8/(1+e^(-y))-0.5*(1-e^(-x))}, v={0.8*e^(-y)/(1+e^(-y))-0.5*(1-e^(-y))}, scale arrows=0.48,
                       every arrow/.append style={-latex}}] (x,y,0);   
            \draw [very thin, draw=gray, fill=gray, opacity=0.24]
               (0.732668,0.732668) rectangle (2.4,2.4) ;
            \draw[red] (0.732668,2.4) -- (0.732668,0);
            \draw[red] (2.4,0.732668) -- (0,0.732668);
          \end{axis}
        \end{tikzpicture}
    }
    \hspace{0.1cm}
    \subfigure [$s=3$] {
        \begin{tikzpicture}[font=\scriptsize]
          \begin{axis}[
              view     = {0}{90},
              domain   = 0:2.4,
              y domain = 0:2.4,
              samples  = 10,
              xmax     = 2.4,
              ymax     = 2.4,
              extra x ticks = {0.732668},
              extra x tick labels={$\underline{x}_1$},
              extra y ticks = {0.732668},
              extra y tick labels={$\underline{x}_2$},
              xlabel   = $x_1$,
              ylabel   = $x_2$,
              xlabel style={yshift=0.2cm},
              ylabel style={yshift=-0.2cm}
            ]
            \addplot3 [blue, quiver={u={0.8*e^(-x)/(e^(-x)+1)-0.5*(1-e^(-x))}, v={0.8/(e^(-x)+1)-0.5*(1-e^(-y))}, scale arrows=0.48,
                       every arrow/.append style={-latex}}] (x,y,0);   
            \draw [very thin, draw=gray, fill=gray, opacity=0.24]
               (0.732668,0.732668) rectangle (2.4,2.4) ;
            \draw[red] (0.732668,2.4) -- (0.732668,0);
            \draw[red] (2.4,0.732668) -- (0,0.732668);
          \end{axis}
        \end{tikzpicture}
    }
    \hspace{0.1cm}
    \subfigure [$s=4$] {
        \begin{tikzpicture}[font=\scriptsize]
          \begin{axis}[
              view     = {0}{90},
              domain   = 0:2.4,
              y domain = 0:2.4,
              samples  = 10,
              xmax     = 2.4,
              ymax     = 2.4,
              extra x ticks = {0.732668},
              extra x tick labels={$\underline{x}_1$},
              extra y ticks = {0.732668},
              extra y tick labels={$\underline{x}_2$},
              xlabel   = $x_1$,
              ylabel   = $x_2$,
              xlabel style={yshift=0.2cm},
              ylabel style={yshift=-0.2cm}
            ]
            \addplot3 [blue, quiver={u={0.8/2-0.5*(1-e^(-x))}, v={0.8/2-0.5*(1-e^(-y))}, scale arrows=0.48,
                       every arrow/.append style={-latex}}] (x,y,0);   
            \draw [very thin, draw=gray, fill=gray, opacity=0.24]
               (0.732668,0.732668) rectangle (2.4,2.4) ;
            \draw[red] (0.732668,2.4) -- (0.732668,0);
            \draw[red] (2.4,0.732668) -- (0,0.732668);
          \end{axis}
        \end{tikzpicture}
    }
    \caption{Illustration of the continuous state process and the invariant set $\mathcal{M}$. The arrows represent the vector field $G$ defined in \eqref{eq_G} for the four states.}
    \label{fig:quiver}
\end{figure}

Second, by ergodicity of the process $\{(S(t),X(t));t>0\}$ where $X(t)=
\begin{bmatrix}
X_1(t) \\ X_2(t)
\end{bmatrix}$, we have for $k\in\{1,2\}$,
$$X_k(t)=X_k(0)+\int_0^t\Big(u_k(\tau)-f_k(\tau)\Big){d}\tau,$$
where $u_k(\tau)$ and $f_k(\tau)$ are inflow and outflow of link $k$ at time $\tau$. Since $\lim_{t\to\infty}\frac1tX_k(0)=0$ and $\lim_{t\to\infty}\frac1tX_k(t)=0$ a.s., then
\begin{align*}
    0
    =\lim_{t\to\infty}\frac1t\Bigg(\int_0^t\Big(u_k(\tau)-f_k(\tau)\Big){d}\tau+X_k(0)-X_k(t)\Bigg)
    =\lim_{t\to\infty}\frac1t\int_0^t\Big(u_k(\tau)-f_k(\tau)\Big){d}\tau \quad\text{a.s.}
\end{align*}
Note that $f_k(\tau)\leq F_k$ for any $\tau\ge0$ and $k\in\{1,2\}$, hence
\begin{align}
  \lim_{t\to\infty}\frac1t\int_0^t u_k(\tau){d}\tau=
  \lim_{t\to\infty}\frac1t\int_0^t f_k(\tau){d}\tau 
  \leq
  \lim_{t\to\infty}\frac1t\int_0^t F_k{d}\tau
  =F_k.  \label{flow}
\end{align}
According to the definition of steady-state probability,
$$\lim_{t\to\infty}\frac1t\int_0^t\mathbb{I}_{S(\tau)=s}{d}\tau=p_s,\quad\text{a.s.}\quad\forall s\in\mathcal{S}.$$
Combining with \eqref{flow}, we obtain
\begin{align*}
    F_1\geq&\lim_{t\to\infty}\frac1t\int_0^t u_1(\tau){d}\tau
    =\lim_{t\to\infty}\frac1t\int_0^t \eta\mu_1(S(\tau), X(\tau)){d}\tau \\
    =&\eta\lim_{t\to\infty}\frac1t\sum_{s=1}^4\int_0^t\mathbb{I}_{S(\tau)=s}\mu_1(S(\tau), X(\tau)){d}\tau \\
    \geq&\eta\lim_{t\to\infty}\frac1t\Big(\int_0^t\mathbb{I}_{S(\tau)=1}0{d}\tau+\int_0^t\mathbb{I}_{S(\tau)=2}\frac{1}{1+e^{-\beta \underline{x_2}}}{d}\tau 
    +\int_0^t\mathbb{I}_{S(\tau)=3}0{d}\tau+\int_0^t\mathbb{I}_{S(\tau)=4}\frac12{d}\tau\Big) \\
    =&\eta\Big(\frac{1}{1+e^{-\beta \underline{x_2}}}\lim_{t\to\infty}\frac1t\int_0^t\mathbb{I}_{S(\tau)=2}{d}\tau
    +\frac12\lim_{t\to\infty}\frac1t\int_0^t\mathbb{I}_{S(\tau)=4}{d}\tau\Big) 
    \\
    =
    &
    \eta\Big(\frac{p_2}{1+e^{-\beta\underline{x_2}}}+\frac{p_4}{2}\Big),
\end{align*}
which gives \eqref{necessary1}. We can prove \eqref{necessary2} in a similar way.

\subsection{Proof of sufficient condition}


Suppose that there exists a vector $\theta\in\mathbb R_{\ge0}^2$ satisfying \eqref{sufficient}.
Then, for the hybrid process $\{(S(t),X(t));t>0\}$, consider the Lyapunov function
\begin{align}
V(s,x)
=\frac12\Big((x_1-\theta_1)_++(x_2-\theta_2)_+\Big)^2
+a_{s}\Big((x_1-\theta_1)_++(x_2-\theta_2)_+\Big)
\label{V}
\end{align}
where $(x_k-\theta_k)_+=\max\{0,x_k-\theta_k\}$, $k=1,2$, and the coefficients $a_{s}$ are given by
\begin{align*}
[a_1, a_2, a_3, a_4]^T= 
{\footnotesize\left[\begin{array}{cccc}
     -\sum\limits_{i\neq1}\lambda_{1i} & \lambda_{12} & \lambda_{13} & \lambda_{14}  \\
     \lambda_{21} & -\sum\limits_{i\neq2}\lambda_{2i} & \lambda_{23} & \lambda_{24}  \\
     \lambda_{31} & \lambda_{32} & -\sum\limits_{i\neq3}\lambda_{3i} & \lambda_{34}\\
     1 & 0 & 0 & 0
\end{array}\right]^{-1} \left[\begin{array}{c}
     \bar G-G(1,\theta)  \\
     \bar G-G(2,\theta)\\
     \bar G-G(3,\theta)\\
     1
\end{array}\right]}
\end{align*}
where $G$ is defined in \eqref{eq_G} and $\bar G=\sum_{s\in\mathcal S}p_sG(s,\theta)$. Based on the ergodicity assumption of the mode switching process, the matrix in the above must be invertible.
This Lyapunov function is valid in that $V(s,x)\to\infty$ as $|x|\to\infty$ for all $s$.
Define
\begin{align}
\mathscr D_s=\max_{k\in\{1,2\}}\Big(\mu_k(s,\theta)-f_k(\theta_k)\Big),
\quad s\in\mathcal S.
\label{Ds}
\end{align}

The Lyapunov function $V$ essentially penalizes the quantity $(x-\theta)_+$, which can be viewed as a ``derived state''.
Apparently, boundedness of $X(t)$ is equivalent to the boundedness of $(X(t)-\theta)_+$
Note that the dynamic equation of the derived state $(x-\theta)_+$ is slightly different from that of $x$:
\begin{align*}
    \frac{d}{dt}(X_k(t)-\theta_k)_+
    =D_k(S(t),X(t)):=
    \begin{cases}
    \mu_k(S(t),X(t))-f_k(X(t) & X_k(t)>\theta_k,\\
    (\mu_k(S(t),X(t))-f_k(X(t))_+ & X_k(t)=\theta_k,\\
    0 & \mbox{otherwise,}
    \end{cases}
    \quad k=1,2.   
\end{align*}

Applying the infinitesimal generator to the Lyapunov function, we obtain
\begin{align}
\mathcal LV(s,x)=
&\sum_{k=1}^2\sum_{j=1}^2D_j(s,x)(x_k-\theta_k)_+
+\sum_{s'\neq s}\Big(\lambda_{s,s'}(a_{s'}-a_{s})\sum_{k=1}^2(x_k-\theta_k)_+\Big)
+\sum_{k=1}^2a_{s,k}D_k(s,x)\nonumber\\
=&\Big(\sum_{k=1}^2D_k(s,x)+\sum_{s'\neq s}\lambda_{s,s'}(a_{s'}-a_{s})\Big)
|(x_k-\theta_k)_+|
+\sum_{k=1}^2a_{s,k}D_k(s,x)
\label{LV}
\end{align}

This proof establishes the stability of the process $\{(S(t),X(t));t>0\}$ by verifying that the Lyapunov function $V$ as defined above satisfies the Foster-Lyapunov drift condition for stability \cite{meyn1993stability}:
\begin{align}
    \mathcal LV(s,x)\le -c|x|+d
    \quad \forall (s,x)\in\mathcal S\times\mathbb R_{\ge0}^2
    \label{drift}
\end{align}
for some $c>0$ and $d<\infty$, where $|x|$ is the one-norm of $x$; this condition will imply \eqref{eq_bounded}.
To proceed, we partition $\mathbb R_{\ge0}^2$, the space of $x$, into two subsets:
$$
\mathcal X_0=\{x:0\le x\le\theta\},\
\mathcal X_1=\mathcal X_0^C;
$$
that is, $\mathcal X_0$ and $\mathcal X_1$ are the complement to each other in the space $\mathbb R_{\ge0}^2$. In the rest of this proof, we first verify \eqref{drift} over $\mathcal X_0$ and then over $\mathcal X_1$.

To verify \eqref{drift} over $\mathcal X_0$, note that $\mu$ and $f$ are bounded functions, so, for any $a_{s,k}$, there exists $d<\infty$ such that
\begin{align}
d_1\ge a_s\sum_{k=1}^2D_k(s,x)
\quad \forall (s,x)\in\mathcal S\times\mathbb R_{\ge0}^2.
\end{align}
In addition, $(x_k-\theta_k)_+=0$, $k=1,2,\ldots,K$ for all $x\in\mathcal X_0$; this and \eqref{LV} imply
\begin{align}
    \mathcal LV(s,x)\le d_1.
    \label{d1}
\end{align}
Furthermore, for any $c>0$, there exists $d_2=c|\theta|$ such that $d_2\ge c|x|$ for all $x\in\mathcal X_0$. Hence, letting $d=d_1+d_2$, we have
\begin{align}
\mathcal LV(s,x)\le -c|x|+d
    \quad \forall (s,x)\in\mathcal S\times\mathcal X_0.
    \label{LVX0}
\end{align}

To verify \eqref{drift} over $\mathcal X_1$, we further decompose $\mathcal X_1$ into the following subsets:
\begin{align*}
    &\mathcal X_1^1=\{x\in\mathcal X_1:x_1\ge\theta_1,x_2<\theta_2\},\\
    &\mathcal X_1^2=\{x\in\mathcal X_1:x_1<\theta_1,x_2\ge\theta_2\},\\
    &\mathcal X_1^3=\{x\in\mathcal X_1:x_1\ge\theta_1,x_2\ge\theta_2\}.
\end{align*}
For each $x\in\mathcal X_1^1$, we have
\begin{align}
    \mathcal LV(s,x)
    &
    =
    \Big(D_1(s,x)+\sum_{s'\neq s}\lambda_{s,s'}(a_{s'}-a_{s})\Big)|(x-\theta)_+| 
    +a_s\sum_{k=1}^2D_k(s,x)\nonumber\\
    &\le \Bigg(\Big(\mu_1(s,x)-f_1(x_1)\Big)+\sum_{s'\neq s}\lambda_{s,s'}(a_{s'}-a_{s})\Bigg)|(x-\theta)_+|
    +d_1
    \nonumber\\
    &
    \le\Big(\mathscr D_s+\sum_{s'\neq s}\lambda_{s,s'}(a_{s'}-a_{s})\Big)|(x-\theta)_+|+d_1
    \label{LV11}
\end{align}
From the definition of $a_s$, we have
$$
\mathscr D_s+\sum_{s'\neq s}\lambda_{s,s'}(a_{s'}-a_{s})=\frac14\sum_{s'\in\mathcal S}p_{s'}\mathscr D_{s'}
$$
The above and \eqref{LV11} imply
\begin{align*}
    \mathcal LV(s,x)\le\frac14\Bigg(\sum_{s'\in\mathcal S}p_{s'}\mathscr D_{s'}\Bigg)|x|+d,
    \quad x\in\mathcal X_1^1.
\end{align*}
Let $c:=-\frac14\sum_{s'\in\mathcal S}p_{s'}\mathscr D_{s'}$. From \eqref{sufficient}, we have $c>0$. Hence, we have
$$
\mathcal LV(s,x)\le-c|x|+d,
\quad\forall (s,x)\in\mathcal X_1^1.
$$
Analogously, we can show
$$
\mathcal LV(s,x)\le-c|x|+d,
\quad\forall (s,x)\in\mathcal X_1^2\cup\mathcal X_1^3,
$$
and hence
$$
\mathcal LV(s,x)\le-c|x|+d,
\quad\forall (s,x)\in\mathcal X_1,
$$
The above and \eqref{LVX0} imply the drift condition \eqref{drift}, which completes the proof.
\section{Resilience analysis}\label{sec:resilience}

In this section, we study the resilience score, i.e. the guaranteed throughput (the supremum of $\eta$ that maintains stability), under various scenarios.
We first consider two symmetric links and focus on the impact of transition rates of the discrete state (Section~\ref{subsec:A}).
Then, we study how the throughput varies with the asymmetry of the links (Section~\ref{subsec:B}).

\subsection{Impact of transition rates}\label{subsec:A}

If the two links are homogeneous in the sense that they have same flow functions $f_1=f_2$, we have the main result of this section as follows:

\begin{prp}\label{thm_resilience}
For the homogeneous network, the resilience score $\eta^*$, i.e. the guaranteed throughput has a lower bound of
\begin{align}
    \eta^*\ge\frac1{1+p_2+p_3}.
\end{align}
\end{prp}

\emph{Proof}:
The lower bound results from the sufficient condition in Theorem~\ref{thm_stability}.

The homogeneity implies that $F_1=F_2=1/2$ and $\theta_1=\theta_2$. Now \eqref{sufficient} means that there exists $\theta_1\in\mathbb R_{\ge0}$ such that
    $$\Big(\frac12(p_1+p_4)+\frac{1}{1+e^{-\beta\theta_1}}(p_2+
    p_3)\Big)\eta < \frac12(1-e^{-\theta_1}),$$
that is,
\begin{align}\label{homo}
    \Big(1+\frac{1-e^{-\beta\theta_1}}{1+e^{-\beta\theta_1}}(p_2+p_3)\Big)\eta < 1-e^{-\theta_1}.
\end{align}

Let $z=e^{-\theta_1}\in(0,1]$, then \eqref{homo} can be expressed as there exists $z\in(0,1]$ such that
    $$\Big(1+\frac{1-z^\beta}{1+z^\beta}(p_2+p_3)\Big)\eta < 1-z, $$
that is, 
\begin{align}\label{z}
    z^{\beta+1}-\Big(1-(1-p_2-p_3)\eta\Big)z^\beta 
    +
    z-\Big(1-(1+p_2+p_3)\eta\Big) < 0.
\end{align}

Let $g(z)$ be the left-hand side of \eqref{z}. Since $g(z)$ is monotonically increasing on (0,1] (proof is provided in Appendices), $\eta$ should satisfy $$g(0)=(1+p_2+p_3)\eta-1<0,$$ or $$\eta < \frac{1}{1+p_2+p_3},$$ which gives the lower bound.

\hfill$\square$

\begin{table}[hbt]
\centering
\caption{Nominal model parameters.}
\begin{tabular}{@{}lcc@{}}
\toprule
Parameter & Notation & Nominal value \\ \midrule
Link 1 capacity        & $F_1$  &  0.5             \\
Link 2 capacity          & $F_2$    &  0.5             \\
Routing sensitivity to congestion & $\beta$ & 1
\\ \bottomrule
\end{tabular}
\label{tab:nominal}
\end{table}

Next, we discuss how characteristics of link failures (specifically, link failure rate and link failure correlation) affect the resilience score. Table \ref{tab:nominal} lists the nominal values considered in this subsection.

\begin{figure}[h]
    \centering
    \begin{tikzpicture}[font=\scriptsize]
    \begin{axis}[
        axis lines = left,
        xlabel = link failure rate $p$,
        ylabel = {lower bound $\underline{\eta^*}$},
    ]
    \addplot [
        domain=0:1, 
        samples=100, 
        color=blue,
        ]
        {1/(1+2*x-2*x^2))};
    \end{axis}
    \end{tikzpicture}
    \qquad
    \begin{tikzpicture}[font=\scriptsize]
    \begin{axis}[
        axis lines = left,
        xlabel = link failure correlation $\rho$,
        ylabel = {lower bound $\underline{\eta^*}$},
        xtick={-0.5,0,0.5},
        xticklabels={-0.5,0,0.5},
        ytick={0.5,0.6,0.7,0.8,0.9,1},
        yticklabels={0.5,0.6,0.7,0.8,0.9,1},
    ]
    \addplot [
        domain=-0.5:0.5, 
        samples=100, 
        color=blue,
        ]
        {1/(1.5-x))};
    \end{axis}
    \end{tikzpicture}
    \caption{Impact of link failure probability $(\rho=0)$ and link failure correlation $(p=0.5)$ on the lower bound of resilience score}
    \label{fig:homo}
\end{figure}
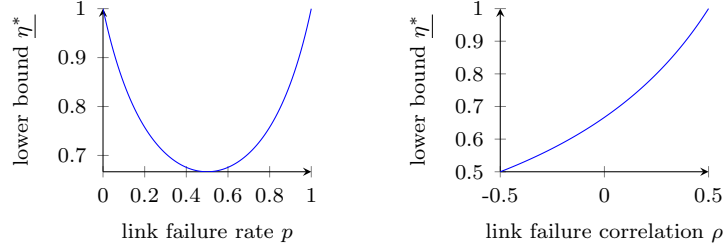

\emph{Link failure rate:} Suppose that the health of each link is independent of the other link. Furthermore, suppose that the failure rates of both links are identical, denoted as $p$, then 
\begin{align*}
&p_2+p_4=p=p_3+p_4,\\
&\underline{\eta^*}=\frac{1}{1+p_2+p_3}=\frac{1}{1+2p(1-p)}.
\end{align*}
When the link failure rate is either 0 or 1, the two-link network becomes open-loop, the lower bound can naturally be 1. The lower bound reaches minimum when the link failure rate is 0.5; see Figure \ref{fig:homo}.

\emph{Link failure correlation:} Suppose that the health of each link is correlated with the other link while the failure rates of both links are still identical. Denote the correlation as $\rho$, then
\begin{align*}
   &\rho = \frac{p_4-(p_2+p_4)(p_3+p_4)}{\sqrt{p_2p_3}} = \frac{p-p_2-p^2}{p},\\
   &\underline{\eta^*}=\frac{1}{1+p_2+p_3}=\frac{1}{1+2p(1-p-\rho)}.
\end{align*}
As the link failure correlation increases from $-p$ to $1-p$, the lower bound increases from $\frac{1}{1+2p}$ to 1. When the failure of the two links are strongly (positively) correlated, the two-link network also turns to be open-loop and hence the lower bound reaches 1; see Figure \ref{fig:homo}.

\subsection{Impact of heterogeneous link capacities}\label{subsec:B}

Now we relax the assumption of symmetric links and allow $F_1\neq F_2$. Without loss of generality, we assume that $F_1\geq F_2$. Instead, we will consider symmetric failure rate, i.e. $p_2=p_3$. The following result links the resilience score to $|F_1-F_2|$, which quantifies the asymmetry of links:

\begin{prp}
\label{prp:hetero}
Suppose that $p_2=p_3$ and $F_1\geq F_2$. Then, the resilience score has a lower bound of
\begin{align}
    \eta^*\ge\min\Big\{\frac{1-(F_1-F_2)}{1-p_1}, \frac{1-p_4(F_1-F_2)}{1+2p_2}\Big\}.
\end{align}
\end{prp}

\emph{Proof}: Let $y=e^{-\theta_1}$, $z=e^{-\theta_2}$, $\rho=\frac{y^\beta-z^\beta}{y^\beta+z^\beta}$.
\eqref{sufficient} implies that there exists $y, z\in (0, 1]$ such that
\begin{align}
&p_1\max\Big\{\frac{\eta y^\beta}{y^\beta+z^\beta}-F_1(1-y), \frac{\eta z^\beta}{y^\beta+z^\beta}-F_2(1-z)\Big\} \nonumber\\
+&p_2\max\Big\{\frac{\eta}{1+z^\beta}-F_1(1-y), \frac{\eta z^\beta}{1+z^\beta}-F_2(1-z)\Big\} \nonumber\\
+&p_3\max\Big\{\frac{\eta y^\beta}{y^\beta+1}-F_1(1-y), \frac{\eta}{y^\beta+1}-F_2(1-z)\Big\} \nonumber\\
+&p_4\max\Big\{\frac\eta2-F_1(1-y), \frac\eta2-F_2(1-z)\Big\}\leq 0
\label{hetero1}
\end{align}

If $\frac{1}{2-p_1}<F_1-F_2\leq1$, when
$$\eta\leq\frac{1-(F_1-F_2)}{1-p_1},$$ there exists $y\leq 1-\frac{\eta+F_2}{F_1}$ such that \eqref{hetero1} holds.

If $0\leq F_1-F_2\leq\frac{1}{2-p_1}$, when $$F_1-F_2\leq\eta\leq\frac{1-(1-p_1-2p_2)(F_1-F_2)}{1+2p_2},$$
there exists $y, z$ satisfying $\rho(F_1-F_2)\geq F_1(1-y)-F_2(1-z)\geq0$ and $\rho<\frac{F_1-F_2}{\eta}$ such that \eqref{hetero1} holds and when
$$\eta<F_1-F_2,$$
there exists $y\leq1-\frac{\eta+F_2}{F_1}$ such that \eqref{hetero1} holds. 

Therefore, $$\eta^*\geq
\begin{cases}
\frac{1-(F_1-F_2)}{1-p_1}, & \frac{1}{2-p_1}<F_1-F_2\leq1 \\
\frac{1-(1-p_1-2p_2)(F_1-F_2)}{1+2p_2}, & 0\leq F_1-F_2\leq\frac{1}{2-p_1}
\end{cases}
$$
The details of the proof are provided in Appendices.

\hfill$\square$

Now we are ready to discuss how link capacity difference affects the resilience score. 

When $F_1=F_2$, the lower bound is $\frac{1}{1+2p_2}$, in consistence with our lower bound in subsection \ref{subsec:A}, and the upper bound is 1 (note that when $\sqrt{2}\max\{p_2, p_3\}+p_4\leq1$, we can derive $$\eta<1$$ from the necessary condition).

As $F_1-F_2$ increases, the lower bound gradually drops and after certain point, it drops faster to 0 while the upper bound remains 1 for a while and then drops to 0. It means that when the difference between two link capacities gets larger, one link starts getting more congested than the other, then the system can be less stable. 

When $F_1\to1$, $F_2\to0$, the network has weak resilience to the sensing faults and the resilience score tends to be zero.
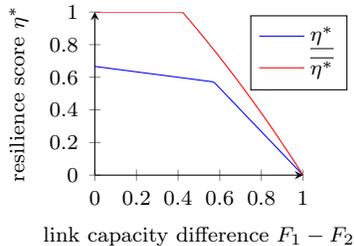
\begin{figure}[h]
    \centering
    \begin{tikzpicture}[font=\scriptsize]
    \begin{axis}[
        axis lines = left,
        xlabel = link capacity difference $F_1-F_2$,
        ylabel = {resilience score $\eta^*$},
        legend style={
                    anchor=north,
                        },
    ]
    \addplot [
        domain=0:1,
        samples=100, 
        color=blue,
        ]
        {min(4/3*(1-x),2/3*(1-0.25*x))};
    \addplot [
        domain=0:1, 
        samples=1000, 
        color=red,
        ]
        {min(1,-2/3*sqrt(3*x^2-6*x+7)-2*x+10/3)};
    \legend{$\underline{\eta^*}$, $\overline{\eta^*}$}
    \end{axis}
    \end{tikzpicture}
    \caption{Impact of link capacity difference on the lower and upper bound of resilience score ($p_1=p_2=p_3=p_4=1/4$)}
    \label{fig:hetero}
\end{figure}
\section{Concluding remarks}\label{sec:concluding}
In this paper, we propose a two-link dynamic flow model with sensing faults to study the stability conditions and guaranteed throughput of the network. Based on this model, we are able to derive lower and upper bounds of the resilience score and analyze the impact of transition rates and heterogeneous link capacities on them. This work can be extended in several directions. First, we can consider a complicated network with $k$ links (not necessarily parallel) rather than a simple two parallel link network. Second, other forms of flow functions can be assumed in the model. Third, the logit model can be replaced with other routing polices. Last, several variations of fault modes can also be discussed.

\bibliographystyle{IEEEtran}
\bibliography{xie20acc}

\section*{Appendices}
\subsection*{The monotonicity of $g(z)$ in subsection \ref{subsec:A}} \label{appen:mono}
The first derivative and the second derivative of $g(z)$ are
    $$g'(z)=(\beta+1)z^\beta-\Big(1-(1-p_2-p_3)\eta\Big)\beta z^{\beta-1}+1,$$
    $$g''(z)=\beta z^{\beta-2}h(z),$$
where $h(z)=(\beta+1)z-\Big(1-(1-p_2-p_3)\eta\Big)(\beta-1)$.

If $0<\beta\le1$, then $h(z)>h(0)=\Big(1-(1-p_2-p_3)\eta\Big)(1-\beta)\geq0$, $g''(z)=\beta z^{\beta-2}h(z)>0$. Hence $g'(z)$ is monotonically increasing on (0,1]. Since $g'(z)>g'(0)=1$, $g(z)$ is also monotonically increasing on (0,1].

If $\beta>1$, let $z_0=\Big(1-(1-p_2-p_3)\eta)\Big)\frac{\beta-1}{\beta+1}$, then $h(z)<0$ on $(0,z_0)$ and $h(z)>0$ on $(z_0,1]$. Since $g''(z)$ has same sign as $h(z)$, $g'(z)\ge g'(z_0)=1>0$. Therefore, $g(z)$ is monotonically increasing on (0,1].

\subsection*{Detailed proof of Proposition \ref{prp:hetero}} \label{appen:detail}
If $\frac{1}{2-p_1}<F_1-F_2\leq1$, then assume $$\eta\leq\frac{1-(F_1-F_2)}{1-p_1},$$
let $y\leq 1-\frac{\eta+F_2}{F_1}$ (note that $\eta\leq\frac{1-(F_1-F_2)}{1-p_1}<F_1-F_2$ means $y$ exists), we have
$$\frac{\eta(1-z^\beta)}{1+z^\beta}<\eta\leq F_1(1-y)-F_2<F_1(1-y)-F_2(1-z).$$
Now \eqref{hetero1} can be expressed as
\begin{align}
    &p_1\Big(\frac{\eta z^\beta}{y^\beta+z^\beta}-F_2(1-z)\Big)+p_2\Big(\frac{\eta z^\beta}{1+z^\beta}-F_2(1-z)\Big)+ \nonumber\\
    &p_3\Big(\frac{\eta}{y^\beta+1}-F_2(1-z)\Big)+p_4\Big(\frac\eta2-F_2(1-z)\Big)\leq0, \label{hetero2}
\end{align}
that is, 
\begin{align*}
    \frac12\Big(1-\frac{y^\beta-z^\beta}{y^\beta+z^\beta} p_1+(\frac{1-y^\beta}{1+y^\beta}-\frac{1-z^\beta}{1+z^\beta})p_2\Big)\eta-F_2(1-z)\leq0.
\end{align*}
Fix $y$ and note that when $z=0$,
\begin{align*}
    \text{LHS}&=\frac12\Big(1-p_1-\frac{2y^\beta}{1+y^\beta}\Big)\eta-F_2 \\
    &<\frac12(1-p_1)\eta-F_2 \\
    &=\frac12-\frac12(F_1-F_2)-F_2 \\
    &=0,
\end{align*}
then intermediate value theorem implies that there exists $z$ such that $\text{LHS}\leq0$.

If $0\leq F_1-F_2\leq\frac{1}{2-p_1}$, first assume $$F_1-F_2\leq\eta\leq\frac{1-(1-p_1-2p_2)(F_1-F_2)}{1+2p_2},$$
let $y, z$ satisfies $\rho(F_1-F_2)\geq F_1(1-y)-F_2(1-z)$ (fix $y$ and note that when $z=0$, $\rho(F_1-F_2)+F_2(1-z)=(F_1-F_2)+F_2=F_1>F_1(1-y)$, intermediate value theorem implies that such $z$ exists) and $F_1(1-y)-F_2(1-z)\geq0$ (let $y\leq 1-\frac{F_2}{F_1}(1-z)$) and $\rho<\frac{F_1-F_2}{\eta}$ (since $\frac{F_1-F_2}{\eta}\leq1$, such $\rho$ exists), we have
$$\rho\eta\geq\rho(F_1-F_2)\geq F_1(1-y)-F_2(1-z).$$
Now \eqref{hetero1} can be expressed as
\begin{align*}
    &p_1\Big(\frac{\eta y^\beta}{y^\beta+z^\beta}-F_1(1-y)\Big)+p_2\Big(\frac{\eta}{1+z^\beta}-F_1(1-y)\Big)+ \\
    &p_3\Big(\frac{\eta}{y^\beta+1}-F_2(1-z)\Big)+p_4\Big(\frac\eta2-F_2(1-z)\Big)\leq0,
\end{align*}
that is,
\begin{align*}
    &\frac12\Big(1+\rho p_1+(\frac{1-z^\beta}{1+z^\beta}+\frac{1-y^\beta}{1+y^\beta})p_2\Big)\eta 
    -
    (p_1+p_2)F_1(1-y)-(1-p_1-p_2)F_2(1-z)\leq0.
\end{align*}
Fix $\rho$ and note that when $z=0$ (and hence $y=0$),
\begin{align*}
    \text{LHS}
    =&\frac12\rho\eta p_1+\frac12(1+2p_2)\eta-(p_1+p_2)(F_1-F_2)-F_2 \\
    <&\frac12(F_1-F_2)p_1+\frac12\Big(1-(1-p_1-2p_2)(F_1-F_2)\Big)
    -(p_1+p_2)(F_1-F_2)-F_2 \\
    =&0,
\end{align*}
then intermediate value theorem implies that there exists $z$ (and $y$) such that $\text{LHS}\leq0$.

Next assume $\eta<F_1-F_2$, let $y\leq1-\frac{\eta+F_2}{F_1}$ (note that $\eta<F_1-F_2$ means such $y$ exists), thus \eqref{hetero1} can also be expressed as \eqref{hetero2}. Note that $\eta<F_1-F_2\leq\frac{1-(F_1-F_2)}{1-p_1}$, we can use similar proof as the case $\frac{1}{2-p_1}<F_1-F_2\leq1$.

\end{document}